\documentclass[%
 reprint,
 amsmath,amssymb,
 aps,
]{revtex4-2}

\usepackage{graphicx}% Include figure files
\usepackage{dcolumn}% Align table columns on decimal point
\usepackage{bm}% bold math
\begin{document}

\preprint{APS/123-QED}

\title{Transition from inhomogeneous to homogeneous broadening at a lasing prethreshold}% Force line breaks with \\

\author{I. S. Pashkevich$^1$}
\author{I. V. Doronin$^{1,2,3}$}
\author{E. S. Andrianov$^{1,2,4}$}
\author{A. A. Zyablovsky$^{1,2,4,5}$}
\email{zyablovskiy@mail.ru}

\affiliation{$^1$Moscow Institute of Physics and Technology, 141700, 9 Institutskiy pereulok, Moscow, Russia}%Lines break automatically or can be forced with \\
\affiliation{$^2$Dukhov Research Institute of Automatics, 127055, 22 Sushchevskaya, Moscow, Russia}%Lines break automatically or can be forced with \\
\affiliation{$^3$Institute of spectroscopy RAS, 108840, 5 Fizicheskaya, Troitsk, Moscow, Russia}%Lines break automatically or can be forced with \\
\affiliation{$^4$Institute for Theoretical and Applied Electromagnetics, 125412, 13 Izhorskaya, Moscow, Russia}%Lines break automatically or can be forced with \\
\affiliation{$^5$Kotelnikov Institute of Radioengineering and Electronics RAS, 125009, 11-7 Mokhovaya, Moscow, Russia}%Lines break automatically or can be forced with \\

\date{\today}% It is always \today, today,

\begin{abstract}
The emission linewidth in active medium emerges due to homogeneous and inhomogeneous broadening. We demonstrate that in lasers with inhomogeneous broadening there is a critical pump rate, above which the special mode forms. This mode consists of locked-in oscillations of cavity mode and of the active particles with different transition frequencies. Below the critical value of the pump rate, the radiation spectrum of the laser has a Gaussian profile, provided that inhomogeneous broadening is dominant. Above the critical value of pump rate, the special mode mostly determines the laser radiation spectrum. As the result, the spectrum attains Lorentz shape characteristic for homogeneous broadening. We demonstrate that the formation of the special mode precedes lasing and that the critical pump rate plays the role of lasing prethreshold. We obtain expressions for the threshold and generation frequency of single-mode laser where both homogeneous and inhomogeneous broadening are present.
\end{abstract}

%Use showkeys class option if keyword
\maketitle

%\tableofcontents
\section{Introduction}
Width of emission spectrum of active medium greatly impacts behavior of any system built on this medium. Linewidth is determined by combination of two types of broadening: homogeneous and inhomogeneous. Homogeneous broadening of emission spectrum occurs due to finite lifetime of excited states of active particles \cite{1,2,3,4,5}. This contribution is dominant when all particles of active medium are identical and therefore share transition frequency. In contrast, inhomogenous broadening occurs due to different transition frequencies of particles which results in collective spectrum being broader than the spectrum of individual particles. This difference in frequencies can originate from various sources. It can take place because particles have different properties, such as varying sizes of quantum dots \cite{6,7,8,9}. Other effects contributing to this type of broadening are different environment near individual particles like inhomogeneous EM mode fields or lattice defects \cite{10,11,12,13}, solute-solvent interaction \cite{14,15,16} and Doppler broadening originating from difference in particle velocities \cite{17,18}.

The presence of inhomogenous broadening leads to emission spectrum with Gaussian shape \cite{16,19,20}. This result can be explained by the central limit theorem \cite{petrov2012sums}, which states that the sum of independent random variables has a Gaussian distribution. In contrast, homogeneous broadening results in a Lorentzian spectrum \cite{20,21}. However, contributions of the two types of broadening can be difficult to distinguish by the emission spectrum alone. In some cases, they are shown to behave similarly, which renders distinction redundant \cite{22,23}.

Despite superficial similarity, the two types of broadening display distinct behavior in a variety of systems. For example, it was discovered that the shape of peaks of the vacuum-field Rabi splitting in strong coupling regime depends only on the homogeneous broadening \cite{24}. In turn, in \cite{25} polaritonic peaks coherence was found to depend crucially on the type of broadening and on the shape of linewidth in general. This result can be applied to various systems, for example semiconductor emitters coupled to optical cavities or ensembles of spins in circuit QED. In \cite{26} materials with inhomogeneous broadening were found to offer lower optical efficiencies than homogeneous counterparts with Nd-doped active medium under diode pumping. Additionally, time-resolved transmission and reflection from an emitter displays oscillations over time in the presence of inhomogeneous broadening \cite{27}.

In this paper, on an example of a single mode laser with dominant inhomogeneous broadening we demonstrate that there is a critical pump rate of the active medium, at which a special mode forms. The special mode includes locked-in oscillations of electromagnetic field in the cavity and of the active particles with different transition frequencies. Rest of eigenmodes consist of polarizations of individual active particles slightly modified by the interaction with the cavity mode. Below the critical pump rate, all eigenmodes give comparable contributions in the laser spectrum. Therefore, the radiation spectrum has Gaussian shape, resulting in the inhomogeneous broadening. Above the critical value, the special mode dominates in the laser spectrum and a single line with a Lorentzian profile appears in the radiation spectrum. So, the inhomogeneous broadening no longer affect the spectrum. At the same time, the lasing threshold and the generation frequency are determined by the linewidth of the inhomogeneous broadening. We show that the formation of the special mode precedes lasing and the critical pump rate plays the role of lasing prethreshold \cite{33}. We believe our findings aid design and study of systems based on active medium with inhomogeneous broadening.

\section{Model of laser with inhomogeneous broadening}
We consider a laser based on a single mode cavity and an active medium consisting of $N$ active particles with two working levels. The role of active particles can be played by, for example, dye molecules \cite{shank1975physics}, quantum dots \cite{zrenner2002coherent,he2019coherently,cartar2017self}, Sr atoms in the magnetic trap \cite{meiser2009prospects,bohnet2012steady}, etc. We designate the resonant frequency of the cavity mode as ${\omega _a}$. The transition frequencies of active particles $\omega _\sigma ^{\left( j \right)}$ have a normal distribution with the expectation value ${\omega _\sigma }$ and the variance $\Delta \omega $ ($\left| {{\omega _\sigma } - {\omega _a}} \right| < \Delta \omega $), which is the case for many inhomogeneously broadened media \cite{16,19,20}.

We use the Maxwell-Bloch equations \cite{29,30} for description of the laser

\begin{equation}
da/dt = \left( { - {\gamma _a} - i{\omega _a}} \right)a - i\Omega \sum\limits_{j = 1}^N {{\sigma _j}}
\label{eq:1}
\end{equation}

\begin{equation}
d{\sigma _j}/dt = \left( { - {\gamma _\sigma } - i\omega _\sigma ^{\left( j \right)}} \right){\sigma _j} + i\,\Omega \,a\,{D_j}
\label{eq:2}
\end{equation}

\begin{equation}
d{D_j}/dt = \left( {{\gamma _P} - {\gamma _D}} \right) - \left( {{\gamma _P} + {\gamma _D}} \right){D_j} + 2i\Omega \left( {a\sigma _j^* + {a^*}{\sigma _j}} \right)
\label{eq:3}
\end{equation}
Here $a$ is the amplitude of electric field in the cavity; ${\sigma _j}$  and ${D_j}$ are the polarization and the population inversion of j-th active particle. $\Omega $ is the coupling strength between the cavity electric field and each of active particles (assumed equal for all particles for simplicity). ${\gamma _a}$ is the relaxation rate of the electric field in the cavity. ${\gamma _D}$ is the relaxation rate of the population inversion of active particles. ${\gamma _P}$ is the pump rate of the active particles. $\gamma_\sigma$ is the dephasing rate of active particles, which determines the width of homogeneous broadening. The index $j$ runs from $1$ to $N$, where $N$ is the total number of active particles.

It is known that there is a threshold pump rate, above which the lasing takes place [29,30]. Below the lasing threshold the stationary solution of Eqns.~(\ref{eq:1})-(\ref{eq:3}) is given as $a = {\sigma _j} = 0$ and $D = {D_0} = \left( {{\gamma _P} - {\gamma _D}} \right)/\left( {{\gamma _P} + {\gamma _D}} \right)$ \cite{29,30}. We perform linear stability analysis of the steady state below the lasing threshold. To this end, we use the equations for small deviations $\delta a$ and $\delta {\sigma _j}$ from the zero stationary state ($a = {\sigma _j} = 0$) \cite{31,32,33}

\begin{equation}
\frac{d}{{dt}}\left( {\begin{array}{*{20}{c}}
  {\delta a} \\ 
  {\delta {\sigma _j}} 
\end{array}} \right) = \left( {\begin{array}{*{20}{c}}
  { - {\gamma _a} - i{\omega _a}}&{ - i\Omega } \\ 
  {i\Omega {D_0}}&{ - {\gamma _\sigma } - i\omega _\sigma ^{\left( j \right)}} 
\end{array}} \right)\left( {\begin{array}{*{20}{c}}
  {\delta a} \\ 
  {\delta {\sigma _j}} 
\end{array}} \right)
\label{eq:4}
\end{equation}
where the index $j$ runs from $1$ to $N$. We calculate the eigenvalues ${\lambda _k}$ and the eigenmodes ${{\mathbf{e}}_k} = {\left( {a,\sigma_1,...,{\sigma _N}} \right)^T}$ of the matrix in the right side of Eq.~(\ref{eq:4}) for different values of ${D_0}$ (i.e. for different pump rates) [Figure~\ref{fig1}].

Hereinafter, for convenience, we consider the eigenfrequencies ${\omega _k} = i{\lambda _k}$ instead of the eigenvalues ${\lambda _k}$. We study the behavior of eigenfrequencies and eigenmodes with changes in the pump rate to determine the generation threshold and changes in the spectrum of the system.

\section{Results}
\begin{figure*}[htbp]
\centering\includegraphics[width=\linewidth]{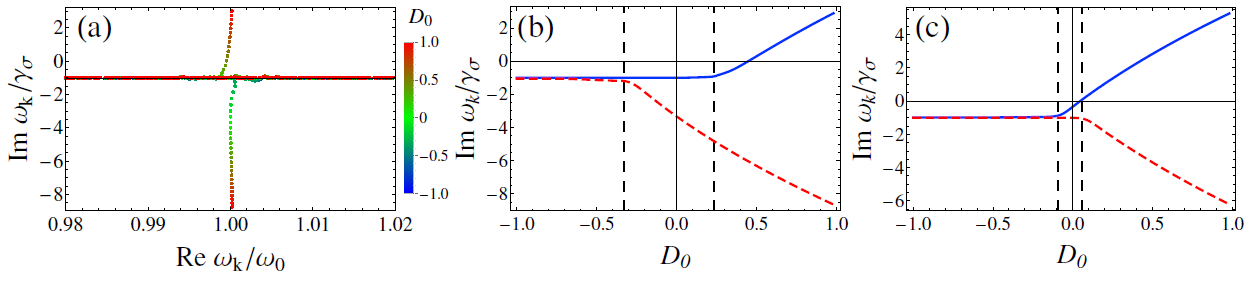}
\caption{(a) Trajectories of eigenfrequencies $\omega_k$ in the complex plane for $D_0$ ranging from $-1$ to $1$. Here ${\gamma _a} = {10^{ - 2}}{\omega _0}$. (b), (c) Dependencies of the imaginary parts of the eigenmodes with the greatest (the dashed red line) and the lowest (the solid blue line) relaxation rate on ${D_0}$ when ${\gamma _a} = {10^{ - 2}}{\omega _0}$ (b), ${\gamma _a} = {10^{ - 3}}{\omega _0}$ (c). All other parameters are the same for three figures: $N = {10^4}$. ${\gamma _\sigma } = 3 \cdot {10^{ - 3}}{\omega _0}$, $\Omega  = 3 \times {10^{ - 4}}\,{\omega _a}$, $\Delta \omega  = 0.05\,{\omega _0}$.}
\label{fig1}
\end{figure*}

Our calculations show that when the pump rate ${\gamma _P}$ tends to zero (${D_0} =  - 1$) the relaxation rates of all eigenmodes tend to ${\gamma _\sigma }$ [Figure~\ref{fig1}a]. This is because active particles give the main contribution to all eigenmodes. An increase in pump rate leads to a change of the relaxation rates of eigenmodes ($\operatorname{Im} \,{\omega _k}$). Notably, there are two critical values of pump rates at which two special eigenmodes form [Figure~\ref{fig1}]. One of these eigenmodes has the greatest relaxation rate among all eigenmodes, and the other special eigenmode has the lowest relaxation rate [Figure~\ref{fig1}a]. When ${\gamma _a} > {\gamma _\sigma }$, the special mode with greatest relaxation rate forms at ${D_0} < 0$, whereas the other special mode forms at ${D_0} > 0$ [Figure~\ref{fig1}b]. The opposite situation takes place when ${\gamma _a} < {\gamma _\sigma }$ [Figure~\ref{fig1}c].

When the pump rate exceeds both of these critical values, the special mode with the greatest relaxation rate has a negligible effect on the system dynamics because of its fast decay. Therefore, we do not consider this mode in further discussion. At the same time, the relaxation rate of the special mode with the lowest relaxation rates further decreases with the pump rate increase, as its eigenfrequency moves up in the complex plane [Figure~\ref{fig1}]. As the result, this special mode exerts more influence on system spectrum. It is this mode which determines behavior of the system above the critical pump rates, and so we will focus on it.

\begin{figure}[htbp]
\centering\includegraphics[width=\linewidth]{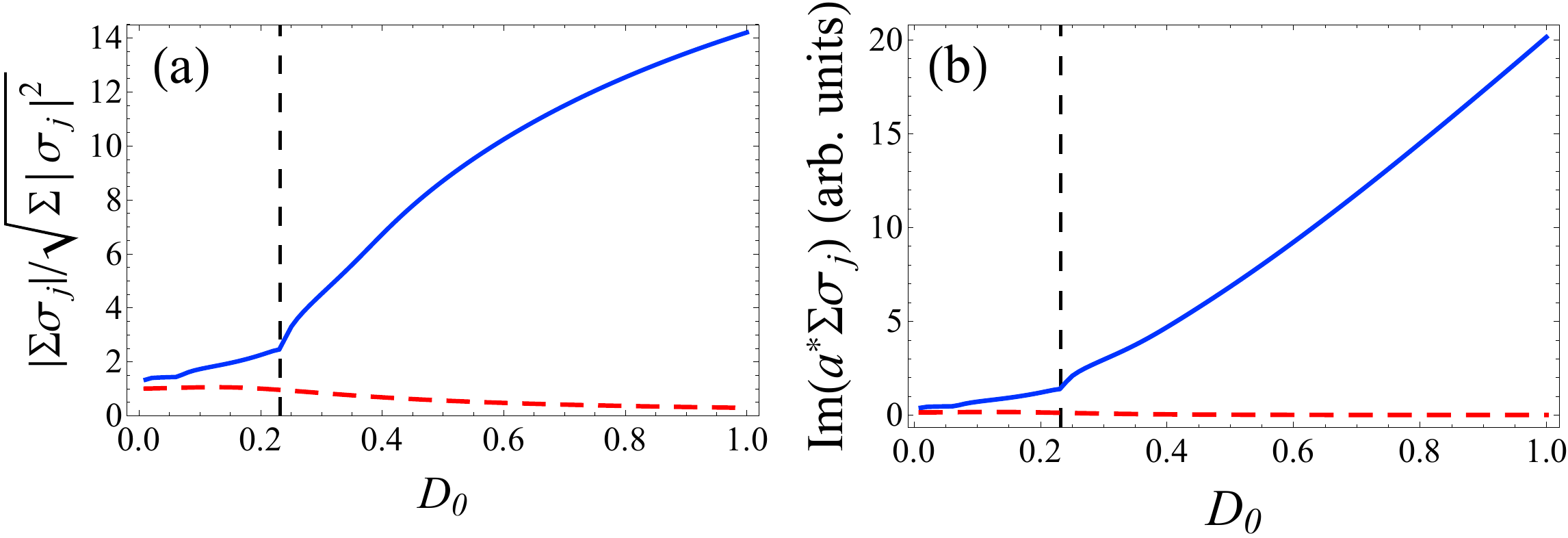}
\caption{(a) Dependence of the modulus of the total polarization value of all active particles, $\left| {\sum\nolimits_j {{\sigma _j}} } \right|$, on ${D_0}$. The total polarization value is normalized to the square root of the sum of the squares of the polarization modules. (b) Dependence of the energy flow from the active particles to the electromagnetic field mode. The blue lines describes the special mode, the red lines describes a typical non-special eigenmode. Here ${\gamma _a} = {10^{ - 3}}{\omega _0}$. All other parameters are the same as for Figure~\ref{fig1}.}
\label{fig2}
\end{figure}

To clarify the mechanism of the special mode formation, we study changes in eigenmodes occurring due to the increase in pump rate. Our calculations show that below both critical values of pump rate, each eigenmode is predominantly associated with oscillations of one of the active particles. That is, for each eigenmode, there is a component which significantly exceeds all other components in absolute value. Above the respective critical value of pump rate, the special eigenmode with the lowest relaxation rate forms. Unlike other modes, this mode has similar absolute values of all components. Consequently, this eigenmode is associated with collective oscillations of the electromagnetic field in the cavity and of the active particles with different transition frequencies. The formation of this special eigenmode leads to phase matching of polarizations of different active particles. As a result, in the special eigenmode the total polarization of all active particles, $\left| {\sum\nolimits_j {{\sigma _j}} } \right|$, sharply increases with the increase in pump rate above the critical pump rate [Figure~\ref{fig2}a]. Moreover, the energy flow from the active medium to the cavity, which is proportional to $ Im \left({a^*}\sum\nolimits_j {{\sigma _j}} \right)$ \cite{34}, also increases [Figure~\ref{fig2}b]. This is accompanied by a fast decrease in relaxation rate of the special mode [Figure~\ref{fig1}]. At the same time, for other eigenmodes the energy flow between the electric field in cavity and the active medium decreases above the critical pump rate [Figure~\ref{fig2}b]. Therefore, these eigenmodes do not experience amplification associated with the interaction between active particles and the cavity mode, thus their relaxation rates remain close to the relaxation rate of free active particles, ${\gamma _\sigma }$.

Thus, above the critical value of pump rate, only the relaxation rate of the special eigenmode decreases with the pump rate. When the relaxation rate of the special mode reaches zero [Figure~\ref{fig1}], the lasing at the special eigenmode begins.
At the same time, all other eigenmodes have non-zero relaxation rates [Figure~\ref{fig1}]. As the result, above the lasing threshold, the special mode dominates the laser spectrum.

Note that the formation of the special mode always precedes lasing. Therefore, the critical pump rate for the formation of a laser mode can be called the lasing prethreshold \cite{33}.

\section{Influence of the formation of the special mode on the lasing threshold and the laser spectrum}

Below the critical value of pump rate all modes contribute equally to the system spectrum. Since these modes share frequencies with free active atoms (only slightly modified by the interaction with the cavity), their collective spectrum forms a Gaussian shape characteristic for inhomogeneous broadening. In contrast, above the critical value of pump rate, the special mode dominates in the emission spectrum. Since the spectrum of a single mode has a Lorentz profile, the prevalence of the special mode in the laser spectrum leads to suppression of the inhomogeneous broadening. As the result, above the critical value of pump rate, the laser spectrum acquires a Lorentz profile.

To quantify the influence of the homogeneous and inhomogeneous broadening on the laser operation, we find the generation threshold and the generation frequency in a laser where both homogeneous and inhomogeneous broadening exist (see Appendix and Section Method in \cite{doronin2021universal}).

\textit{In the resonant case}, $\omega_a = \omega_\sigma$, the generation frequency, $\omega_g$, is equal to $\omega_a$ and the generation threshold is given by the following expression:

\begin{equation}
{D_0}={D_{th}} = \frac{{{\gamma _a}\Delta \omega }}{{N{\Omega ^2}}}\sqrt {\frac{2}{\pi }} \frac{{\exp \left( { - \gamma _\sigma ^2/2\Delta {\omega ^2}} \right)}}{{{\text{erfc}}\left( {{\gamma _\sigma }/\sqrt 2 \Delta \omega } \right)}}
\label{eq:5}
\end{equation}
Here ${\text{erfc}}\left( x \right) = 1 - \frac{2}{\sqrt{\pi} }\int_0^x {\exp ( - {t^2})dt}$ is the complementary error function \cite{abramowitz1968handbook} and $D_0$ is determined by the pump rate as ${D_0} = \left( {{\gamma _P} - {\gamma _D}} \right)/\left( {{\gamma _P} + {\gamma _D}} \right)$.

Two notable limiting cases can be obtained from Eq.~(\ref{eq:5}). For ${\gamma _\sigma } \gg \Delta \omega $ Eq.~(\ref{eq:5}) reduces to the following equation:

\begin{equation}
{D_{th}} = \frac{{{\gamma _\sigma }{\gamma _a}}}{{N\,{\Omega ^2}}}
\label{eq:6}
\end{equation}  
Here we use the fact that, in the limit when $x$ tends to infinity, ${\text{erfc}}(x) \approx \frac{{\exp ( - {x^2})}}{{x\sqrt \pi  }}$ \cite{abramowitz1968handbook}. 
Eq.~(\ref{eq:6}) coincides with the well-known expression for the lasing threshold in the case where only homogeneous broadening is present \cite{29}. 

For ${\gamma _\sigma } \ll \Delta \omega $ the complementary error function is close to unity \cite{abramowitz1968handbook} and Eq.~(\ref{eq:5}) reduces to

\begin{equation}
{D_{th}} = \frac{{{\gamma _a}\Delta \omega }}{{N{\Omega ^2}}}\sqrt {\frac{2}{\pi }}
\label{eq:7}
\end{equation}

\textit{In the non-resonant case}, ${\omega _a} \ne {\omega _\sigma }$, the generation frequency, ${\omega _g}$, depends non-monotonically on the ratio of the width of the homogeneous broadening to the width of the inhomogeneous broadening [Figure~\ref{fig3}] (see Appendix for details of calculations). If we assume $\Delta \omega  \gg {\gamma _\sigma }$ and $\Delta \omega  \gg \left| {{\omega _a} - {\omega _\sigma }} \right|$, the generation frequency is found to be (see Appendix):

\begin{equation}
{\omega _g} = \frac{{\sqrt {\frac{\pi }{2}} \Delta \omega \,{\omega _a} + {\gamma _a}{\omega _\sigma }}}{{\sqrt {\frac{\pi }{2}} \Delta \omega  + {\gamma _a}}}
\label{eq:8}
\end{equation}
At the same time, in the leading order in $\frac{\gamma_\sigma}{\Delta \omega}$, the generation threshold is still determined by Eq.~(\ref{eq:7}) even in the non-resonant case. This is because we assume $\Delta \omega  \gg \left| {{\omega _a} - {\omega _\sigma }} \right|$, which means that the active medium linewidth is much greater than the detuning, and so the detuning becomes irrelevant for threshold condition. 

In the opposite case, when $\Delta \omega  \ll {\gamma _\sigma }$, the  generation frequency in a laser is given by the following exression:

\begin{equation}
{\omega _g} = \frac{{{\gamma _\sigma }{\omega _a} + {\gamma _a}{\omega _\sigma }}}{{{\gamma _\sigma } + {\gamma _a}}}
\label{eq:9}
\end{equation}
This expression coincides with the well-known formula for the generation frequency in a laser with homogeneous broadening \cite{29}.

\begin{figure}[htbp]
\centering\includegraphics[width=0.7\linewidth]{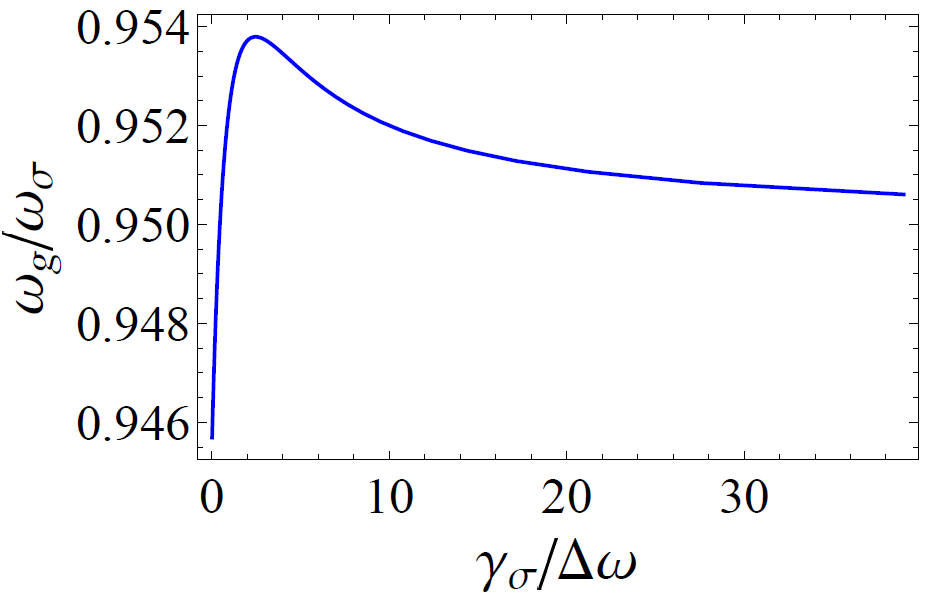}
\caption{Dependence of the generation frequency of the laser, $\omega_g$, on ratio between homogeneous width, $\gamma_\sigma$, and inhomogeneous width, $\Delta \omega$, when $\omega_a \ne \omega_\sigma$. The generation frequency is calculated by using Eq.~(\ref{eq:8A}). Here $\gamma_a =0.1 \omega_\sigma$, ${\gamma _\sigma } + \Delta \omega  = 0.1 \omega_\sigma$, $\omega_a =0.9 \omega _\sigma$.}
\label{fig3}
\end{figure}

Comparing Eq.~(\ref{eq:6}) and Eq.~(\ref{eq:7}), we find that the expressions for the generation thresholds are the same up to the replacement of $\gamma_\sigma$ by $\sqrt {\frac{2}{\pi }} \Delta \omega$. Therefore, an active medium with dominant inhomogeneous broadening has a slightly lower lasing threshold than an active medium with the same spectral width, but originating from homogeneous broadening. At the same time, from Eq.~(\ref{eq:8}) and Eq.~(\ref{eq:9}) it is clear that within the given assumptions the generation frequency are the same up to the replacement of $\gamma_\sigma$ by $\sqrt {\frac{\pi}{2}} \Delta \omega$. Therefore, in an active medium with dominant inhomogeneous broadening the generation frequency is closer to the cavity frequency than in an active medium with the same spectral width, but originating from homogeneous broadening. Thus, homogeneous and inhomogeneous broadening have slightly different effects on the generation threshold and the generation frequency. However, same expressions can be used with reasonable accuracy for systems with dominant homogeneous broadening and dominant inhomogeneous broadening.

\section{Conclusion}
We have studied the effect of inhomogeneous broadening on behavior of single mode lasers. We show that there is a critical value of pump rate, at which a special mode forms. This special mode consists of collective oscillations of the electromagnetic field in the cavity mode and of the active particles with different transition frequencies. Despite inhomogeneous broadening, in this mode the oscillations of all active particles occur at a single shared frequency. Further increase in the pump rate leads to a decrease in the relaxation rate of the special eigenmode, eventually resulting in lasing. Above the lasing threshold, the special mode dominates the laser spectrum. We demonstrate that due to the frequency matching of contributions of active particle's polarizations in this eigenmode, the inhomogeneous broadening is effectively replaced by a homogeneous broadening, and a single mode laser can be described in terms of homogeneously broadened medium.

In terms of behavior, two key features have been identified. First, the inhomogeneously broadened emission slowly transitions into a single homogeneously broadened mode with a Lorentz spectrum as the laser approaches lasing threshold. Therefore, above the threshold inhomogeneous and homogeneous broadening in the active medium are indistinguishable in the spectrum of a single mode laser. Second, inhomogeneous broadening affects the lasing threshold and the generation frequency in a similar way compared to homogeneous broadening, replacing active particle homogeneous linewidth in both formulas when the inhomogeneous broadening is dominant. However, the lasing threshold for inhomogeneously broadened active medium is slightly lower than the one for homogeneously broadened medium. We attribute this advantage to the fact that Gaussian shape decays faster away from the peak compared to Lorentz shape, therefore, for the two distributions with the same full width at half maximum Gaussian distribution is effectively narrower. This finding is valuable for design of low-threshold lasers, where inhomogeneous broadening may result in lower threshold pump rate than expected.

\section*{Appendix}

To derive expression for the lasing threshold we find condition that enables non-trivial stationary solution in Eqns.~(\ref{eq:1}), (\ref{eq:2}) for constant population inversion, ${D_j} = {D_{th}},j = 1,...$. We assume that ${\omega _a} = {\omega _\sigma }$ and look for a solution in the form $a = {a_0}{e^{ - i{\omega _g}t}}$, ${\sigma _j} = {\sigma _{0j}}{e^{ - i{\omega _g}t}},j = 1,...$ (${\omega _g}$ is yet unknown generation frequency). Substituting ${a_0}$ from Eq.~(\ref{eq:1}) into Eq.~(\ref{eq:2}) we arrive at the condition:

\begin{equation}
{\sigma _{0j}} = \frac{{{D_{th}}{\Omega ^2}\sum\nolimits_k {{\sigma _{0k}}} }}{{\left( {{\gamma _a} + i\left( {{\omega _a} - {\omega _g}} \right)} \right)\left( {{\gamma _\sigma } + i\,(\omega _\sigma ^{(j)} - \,{\omega _g})} \right)}},\,\,j = 1,...
\label{eq:1A}
\end{equation}
    
We proceed to sum Eq.~(\ref{eq:1A}) over $j$ (over all active particles). Existence of a nontrivial solution then demands that 

\begin{equation}
\frac{{{\gamma _a} + i\left( {{\omega _a} - {\omega _g}} \right)}}{{{\Omega ^2}}} = {D_{th}}\sum\nolimits_j {\frac{1}{{\left( {{\gamma _\sigma } + i\,(\omega _\sigma ^{(j)} - \,{\omega _g})} \right)}}}
\label{eq:2A}
\end{equation}
  
Imaginary part of Eq.~(\ref{eq:2A}) results in ${\omega _g} = {\omega _a}$ for any symmetrical distribution of $\omega _\sigma ^{(j)}$ centered around $\omega_a$ (as expected in resonant case). The real part of Eq.~(\ref{eq:2A}) yields:

\begin{equation}
\frac{{{\gamma _a}}}{{{\Omega ^2}}} = {\gamma _\sigma }{D_{th}}\sum\nolimits_j {\frac{1}{{\gamma _\sigma ^2 + \,{{(\omega _\sigma ^{(j)} - \,{\omega _a})}^2}}}}
\label{eq:3A}
\end{equation}
  
For a sufficiently large number of active particles with Gaussian distribution (a typical case for inhomogeneously broadened medium \cite{16,19,20}) and ${\omega _a} \gg \Delta \omega ,{\gamma _\sigma },{\gamma _a}$ we can transform the sum in Eq.~(\ref{eq:3A}) into an integral over frequencies to obtain:

\begin{equation}
\frac{{{\gamma _a}}}{{{\Omega ^2}}} = {\gamma _\sigma }{D_{th}}\int_{ - \infty }^{ + \infty } {\frac{{Ndx}}{{\Delta \omega }}\frac{1}{{\gamma _\sigma ^2 + \,{x^2}}}\frac{{\exp \left( { - {x^2}/2\Delta {\omega ^2}} \right)}}{{\sqrt {2\pi } }}}
\label{eq:4A}
\end{equation}
  
The integral in Eq.~(\ref{eq:4A}) can be obtained analytically \cite{abramowitz1968handbook}, which results into:

\begin{equation}
{D_{th}} = \frac{{{\gamma _a}\Delta \omega }}{{N{\Omega ^2}}}\sqrt {\frac{2}{\pi }} \frac{{\exp \left( { - \gamma _\sigma ^2/2\Delta {\omega ^2}} \right)}}{{{\text{erfc}}\left( {{\gamma _\sigma }/\sqrt 2 \Delta \omega } \right)}}
\label{eq:5A}
\end{equation}
where ${\text{erfc}}\left( x \right) = 1 - \frac{2}{\sqrt{\pi} }\int_0^x {\exp ( - {t^2})dt}$ is the complementary error function \cite{abramowitz1968handbook}.

Now we derive expressions for the non-resonant case. First, we need to identify the generation frequency. We divide the real part of Eq.~(\ref{eq:2A}) by the imaginary part of Eq.~(\ref{eq:2A}) to exclude ${D_{th}}$ and obtain relation for the generation frequency ${\omega _g}$:

\begin{equation}
\frac{{{\gamma _a}}}{{{\omega _a} - {\omega _g}}} =  - \frac{{\sum\nolimits_j {\frac{{{\gamma _\sigma }}}{{\gamma _\sigma ^2 + \,{{(\omega _\sigma ^{(j)} - \,{\omega _g})}^2}}}} }}{{\sum\nolimits_j {\frac{{\omega _\sigma ^{(j)} - \,{\omega _g}}}{{\gamma _\sigma ^2 + \,{{(\omega _\sigma ^{(j)} - \,{\omega _g})}^2}}}} }}
\label{eq:6A}
\end{equation}
   
These sums are then transformed into integrals in the same way as the transition from Eq.~(\ref{eq:5A}) to Eq.~(\ref{eq:6A}). We will focus on finding the right side of Eq.~(\ref{eq:7A})

\begin{equation}
\frac{{{\gamma _a}}}{{{\omega _a} - {\omega _g}}} = \frac{{\int_{ - \infty }^\infty  {\frac{{{\gamma _\sigma }\exp ( - {{(x - {\omega _\sigma })}^2}/2\Delta {\omega ^2})}}{{\gamma _\sigma ^2 + \,{{(x - \,{\omega _g})}^2}}}dx} }}{{\int_{ - \infty }^\infty  {\frac{{({\omega _g} - x)\exp ( - {{(x - {\omega _\sigma })}^2}/2\Delta {\omega ^2})}}{{\gamma _\sigma ^2 + \,{{(x - \,{\omega _g})}^2}}}dx} }}
\label{eq:7A}
\end{equation}

Note that ${\omega _\sigma }$ is the center of $\omega _\sigma ^{(j)}$ distribution. Integrals can be calculated as \cite{abramowitz1968handbook}

\begin{equation}
\begin{gathered}
  \frac{{{\gamma _a}}}{{{\omega _a} - {\omega _g}}} =  \hfill \\
  i\frac{{{\text{erfc}}\frac{{{\gamma _\sigma } - i({\omega _g} - {\omega _\sigma })}}{{\sqrt 2 \Delta \omega }} + \exp \frac{{2i{\gamma _\sigma }({\omega _g} - {\omega _\sigma })}}{{\Delta {\omega ^2}}}{\text{erfc}}\frac{{{\gamma _\sigma } + i({\omega _g} - {\omega _\sigma })}}{{\sqrt 2 \Delta \omega }}}}{{{\text{erfc}}\frac{{{\gamma _\sigma } - i({\omega _g} - {\omega _\sigma })}}{{\sqrt 2 \Delta \omega }} - \exp \frac{{2i{\gamma _\sigma }({\omega _g} - {\omega _\sigma })}}{{\Delta {\omega ^2}}}{\text{erfc}}\frac{{{\gamma _\sigma } + i({\omega _g} - {\omega _\sigma })}}{{\sqrt 2 \Delta \omega }}}} \hfill \\ 
\end{gathered}
\label{eq:8A}
\end{equation}

In general, Eq.~(\ref{eq:8A}) can be solved numerically to obtain ${\omega _g}$, however, a useful limit can be obtained if we assume that the inhomogeneous broadening is larger than the homogeneous broadening, $\Delta \omega  \gg {\gamma _\sigma }$, and the frequency detuning, $\Delta \omega  \gg \left| {{\omega _g} - {\omega _\sigma }} \right|$ (the latter is satisfied if, e.g., $\Delta \omega  \gg \left| {{\omega _a} - {\omega _\sigma }} \right|$, since the generation frequency ${\omega _g}$ lies between ${\omega _a}$ and ${\omega _\sigma }$). In this case the complementary error functions and exponents in  Eq.~(\ref{eq:8A}) are estimated as two leading orders of their Taylor series, which results in

\begin{equation}
\frac{{{\gamma _a}}}{{{\omega _a} - {\omega _g}}} = \sqrt {\frac{\pi }{2}} \frac{{\Delta \omega }}{{{\omega _g} - {\omega _\sigma }}}
\label{eq:9A}
\end{equation}

After trivial algebra we obtain

\begin{equation}
{\omega _g} = \frac{{\sqrt {\pi /2} \Delta \omega {\omega _a} + {\gamma _a}{\omega _\sigma }}}{{\sqrt {\pi /2} \Delta \omega  + {\gamma _a}}}
\label{eq:10A}
\end{equation}

Inserting Eq.~(\ref{eq:10A}) into the real part of Eq.~(\ref{eq:2A}) results in the same threshold condition (see Eq.~(\ref{eq:7})). The reason why detuning between electromagnetic field and active medium does not affect the threshold in our derivation is because we assume $\Delta \omega  \gg \left| {{\omega _g} - {\omega _\sigma }} \right|$ to obtain Eq.~(\ref{eq:10A}), i.e. the active medium linewidth is much greater than the detuning, and thus the latter does not play a notable role.

\section*{Acknowledgement}
The study was financially supported by a Grant from Russian Science Foundation (Project No. 22-72-00026). I.V.D. thanks foundation for the advancement of theoretical physics and mathematics “Basis”.

\nocite{*}

\bibliography{apssamp}% Produces the bibliography via BibTeX.

\providecommand{\noopsort}[1]{}\providecommand{\singleletter}[1]{#1}%
\begin{thebibliography}{43}%
\makeatletter
\providecommand \@ifxundefined [1]{%
 \@ifx{#1\undefined}
}%
\providecommand \@ifnum [1]{%
 \ifnum #1\expandafter \@firstoftwo
 \else \expandafter \@secondoftwo
 \fi
}%
\providecommand \@ifx [1]{%
 \ifx #1\expandafter \@firstoftwo
 \else \expandafter \@secondoftwo
 \fi
}%
\providecommand \natexlab [1]{#1}%
\providecommand \enquote  [1]{``#1''}%
\providecommand \bibnamefont  [1]{#1}%
\providecommand \bibfnamefont [1]{#1}%
\providecommand \citenamefont [1]{#1}%
\providecommand \href@noop [0]{\@secondoftwo}%
\providecommand \href [0]{\begingroup \@sanitize@url \@href}%
\providecommand \@href[1]{\@@startlink{#1}\@@href}%
\providecommand \@@href[1]{\endgroup#1\@@endlink}%
\providecommand \@sanitize@url [0]{\catcode `\\12\catcode `\$12\catcode `\&12\catcode `\#12\catcode `\^12\catcode `\_12\catcode `\%12\relax}%
\providecommand \@@startlink[1]{}%
\providecommand \@@endlink[0]{}%
\providecommand \url  [0]{\begingroup\@sanitize@url \@url }%
\providecommand \@url [1]{\endgroup\@href {#1}{\urlprefix }}%
\providecommand \urlprefix  [0]{URL }%
\providecommand \Eprint [0]{\href }%
\providecommand \doibase [0]{https://doi.org/}%
\providecommand \selectlanguage [0]{\@gobble}%
\providecommand \bibinfo  [0]{\@secondoftwo}%
\providecommand \bibfield  [0]{\@secondoftwo}%
\providecommand \translation [1]{[#1]}%
\providecommand \BibitemOpen [0]{}%
\providecommand \bibitemStop [0]{}%
\providecommand \bibitemNoStop [0]{.\EOS\space}%
\providecommand \EOS [0]{\spacefactor3000\relax}%
\providecommand \BibitemShut  [1]{\csname bibitem#1\endcsname}%
\let\auto@bib@innerbib\@empty
%</preamble>
\bibitem [{\citenamefont {Wehrenfennig}\ \emph {et~al.}(2014)\citenamefont {Wehrenfennig}, \citenamefont {Liu}, \citenamefont {Snaith}, \citenamefont {Johnston},\ and\ \citenamefont {Herz}}]{1}%
  \BibitemOpen
  \bibfield  {author} {\bibinfo {author} {\bibfnamefont {C.}~\bibnamefont {Wehrenfennig}}, \bibinfo {author} {\bibfnamefont {M.}~\bibnamefont {Liu}}, \bibinfo {author} {\bibfnamefont {H.~J.}\ \bibnamefont {Snaith}}, \bibinfo {author} {\bibfnamefont {M.~B.}\ \bibnamefont {Johnston}},\ and\ \bibinfo {author} {\bibfnamefont {L.~M.}\ \bibnamefont {Herz}},\ }\bibfield  {title} {\bibinfo {title} {Homogeneous emission line broadening in the organo lead halide perovskite ch3nh3pbi3–x cl x},\ }\href@noop {} {\bibfield  {journal} {\bibinfo  {journal} {J. Phys. Chem. Lett.}\ }\textbf {\bibinfo {volume} {5}},\ \bibinfo {pages} {1300} (\bibinfo {year} {2014})}\BibitemShut {NoStop}%
\bibitem [{\citenamefont {Bartholomew}\ \emph {et~al.}(2017)\citenamefont {Bartholomew}, \citenamefont {de~Oliveira~Lima}, \citenamefont {Ferrier},\ and\ \citenamefont {Goldner}}]{2}%
  \BibitemOpen
  \bibfield  {author} {\bibinfo {author} {\bibfnamefont {J.~G.}\ \bibnamefont {Bartholomew}}, \bibinfo {author} {\bibfnamefont {K.}~\bibnamefont {de~Oliveira~Lima}}, \bibinfo {author} {\bibfnamefont {A.}~\bibnamefont {Ferrier}},\ and\ \bibinfo {author} {\bibfnamefont {P.}~\bibnamefont {Goldner}},\ }\bibfield  {title} {\bibinfo {title} {Optical line width broadening mechanisms at the 10 khz level in eu3+: Y2o3 nanoparticles},\ }\href@noop {} {\bibfield  {journal} {\bibinfo  {journal} {Nano Lett.}\ }\textbf {\bibinfo {volume} {17}},\ \bibinfo {pages} {778} (\bibinfo {year} {2017})}\BibitemShut {NoStop}%
\bibitem [{\citenamefont {Sarkar}\ \emph {et~al.}(2010)\citenamefont {Sarkar}, \citenamefont {Ahuja}, \citenamefont {Vasos},\ and\ \citenamefont {Bodenhausen}}]{3}%
  \BibitemOpen
  \bibfield  {author} {\bibinfo {author} {\bibfnamefont {R.}~\bibnamefont {Sarkar}}, \bibinfo {author} {\bibfnamefont {P.}~\bibnamefont {Ahuja}}, \bibinfo {author} {\bibfnamefont {P.~R.}\ \bibnamefont {Vasos}},\ and\ \bibinfo {author} {\bibfnamefont {G.}~\bibnamefont {Bodenhausen}},\ }\bibfield  {title} {\bibinfo {title} {Long-lived coherences for homogeneous line narrowing in spectroscopy},\ }\href@noop {} {\bibfield  {journal} {\bibinfo  {journal} {Phys. Rev. Lett.}\ }\textbf {\bibinfo {volume} {104}},\ \bibinfo {pages} {053001} (\bibinfo {year} {2010})}\BibitemShut {NoStop}%
\bibitem [{\citenamefont {Moody}\ \emph {et~al.}(2015)\citenamefont {Moody}, \citenamefont {Kavir~Dass}, \citenamefont {Hao}, \citenamefont {Chen}, \citenamefont {Li}, \citenamefont {Singh}, \citenamefont {Tran}, \citenamefont {Clark}, \citenamefont {Xu},\ and\ \citenamefont {Berghäuser}}]{4}%
  \BibitemOpen
  \bibfield  {author} {\bibinfo {author} {\bibfnamefont {G.}~\bibnamefont {Moody}}, \bibinfo {author} {\bibfnamefont {C.}~\bibnamefont {Kavir~Dass}}, \bibinfo {author} {\bibfnamefont {K.}~\bibnamefont {Hao}}, \bibinfo {author} {\bibfnamefont {C.-H.}\ \bibnamefont {Chen}}, \bibinfo {author} {\bibfnamefont {L.-J.}\ \bibnamefont {Li}}, \bibinfo {author} {\bibfnamefont {A.}~\bibnamefont {Singh}}, \bibinfo {author} {\bibfnamefont {K.}~\bibnamefont {Tran}}, \bibinfo {author} {\bibfnamefont {G.}~\bibnamefont {Clark}}, \bibinfo {author} {\bibfnamefont {X.}~\bibnamefont {Xu}},\ and\ \bibinfo {author} {\bibfnamefont {G.}~\bibnamefont {Berghäuser}},\ }\bibfield  {title} {\bibinfo {title} {Intrinsic homogeneous linewidth and broadening mechanisms of excitons in monolayer transition metal dichalcogenides},\ }\href@noop {} {\bibfield  {journal} {\bibinfo  {journal} {Nat. Commun.}\ }\textbf {\bibinfo {volume} {6}},\ \bibinfo {pages} {8315} (\bibinfo {year} {2015})}\BibitemShut {NoStop}%
\bibitem [{\citenamefont {Cadiz}\ \emph {et~al.}(2017)\citenamefont {Cadiz}, \citenamefont {Courtade}, \citenamefont {Robert}, \citenamefont {Wang}, \citenamefont {Shen}, \citenamefont {Cai}, \citenamefont {Taniguchi}, \citenamefont {Watanabe}, \citenamefont {Carrere},\ and\ \citenamefont {Lagarde}}]{5}%
  \BibitemOpen
  \bibfield  {author} {\bibinfo {author} {\bibfnamefont {F.}~\bibnamefont {Cadiz}}, \bibinfo {author} {\bibfnamefont {E.}~\bibnamefont {Courtade}}, \bibinfo {author} {\bibfnamefont {C.}~\bibnamefont {Robert}}, \bibinfo {author} {\bibfnamefont {G.}~\bibnamefont {Wang}}, \bibinfo {author} {\bibfnamefont {Y.}~\bibnamefont {Shen}}, \bibinfo {author} {\bibfnamefont {H.}~\bibnamefont {Cai}}, \bibinfo {author} {\bibfnamefont {T.}~\bibnamefont {Taniguchi}}, \bibinfo {author} {\bibfnamefont {K.}~\bibnamefont {Watanabe}}, \bibinfo {author} {\bibfnamefont {H.}~\bibnamefont {Carrere}},\ and\ \bibinfo {author} {\bibfnamefont {D.}~\bibnamefont {Lagarde}},\ }\bibfield  {title} {\bibinfo {title} {Excitonic linewidth approaching the homogeneous limit in mos 2-based van der waals heterostructures},\ }\href@noop {} {\bibfield  {journal} {\bibinfo  {journal} {Phys. Rev. X}\ }\textbf {\bibinfo {volume} {7}},\ \bibinfo {pages} {021026} (\bibinfo {year} {2017})}\BibitemShut {NoStop}%
\bibitem [{\citenamefont {Rafailov}\ \emph {et~al.}(2007)\citenamefont {Rafailov}, \citenamefont {Cataluna},\ and\ \citenamefont {Sibbett}}]{6}%
  \BibitemOpen
  \bibfield  {author} {\bibinfo {author} {\bibfnamefont {E.~U.}\ \bibnamefont {Rafailov}}, \bibinfo {author} {\bibfnamefont {M.~A.}\ \bibnamefont {Cataluna}},\ and\ \bibinfo {author} {\bibfnamefont {W.}~\bibnamefont {Sibbett}},\ }\bibfield  {title} {\bibinfo {title} {Mode-locked quantum-dot lasers},\ }\href@noop {} {\bibfield  {journal} {\bibinfo  {journal} {Nature Photon.}\ }\textbf {\bibinfo {volume} {1}},\ \bibinfo {pages} {395} (\bibinfo {year} {2007})}\BibitemShut {NoStop}%
\bibitem [{\citenamefont {Salvador}\ \emph {et~al.}(2003)\citenamefont {Salvador}, \citenamefont {Hines},\ and\ \citenamefont {Scholes}}]{7}%
  \BibitemOpen
  \bibfield  {author} {\bibinfo {author} {\bibfnamefont {M.~R.}\ \bibnamefont {Salvador}}, \bibinfo {author} {\bibfnamefont {M.~A.}\ \bibnamefont {Hines}},\ and\ \bibinfo {author} {\bibfnamefont {G.~D.}\ \bibnamefont {Scholes}},\ }\bibfield  {title} {\bibinfo {title} {Exciton–bath coupling and inhomogeneous broadening in the optical spectroscopy of semiconductor quantum dots},\ }\href@noop {} {\bibfield  {journal} {\bibinfo  {journal} {J. Chem. Phys.}\ }\textbf {\bibinfo {volume} {118}},\ \bibinfo {pages} {9380} (\bibinfo {year} {2003})}\BibitemShut {NoStop}%
\bibitem [{\citenamefont {Savchenko}\ and\ \citenamefont {Weinstein}(2019)}]{8}%
  \BibitemOpen
  \bibfield  {author} {\bibinfo {author} {\bibfnamefont {S.~S.}\ \bibnamefont {Savchenko}}\ and\ \bibinfo {author} {\bibfnamefont {I.~A.}\ \bibnamefont {Weinstein}},\ }\bibfield  {title} {\bibinfo {title} {Inhomogeneous broadening of the exciton band in optical absorption spectra of inp/zns nanocrystals},\ }\href@noop {} {\bibfield  {journal} {\bibinfo  {journal} {Nanomater.}\ }\textbf {\bibinfo {volume} {9}},\ \bibinfo {pages} {716} (\bibinfo {year} {2019})}\BibitemShut {NoStop}%
\bibitem [{\citenamefont {Van Der~Bok}\ \emph {et~al.}(2020)\citenamefont {Van Der~Bok}, \citenamefont {Dekker}, \citenamefont {Peerlings}, \citenamefont {Salzmann},\ and\ \citenamefont {Meijerink}}]{9}%
  \BibitemOpen
  \bibfield  {author} {\bibinfo {author} {\bibfnamefont {J.~C.}\ \bibnamefont {Van Der~Bok}}, \bibinfo {author} {\bibfnamefont {D.~M.}\ \bibnamefont {Dekker}}, \bibinfo {author} {\bibfnamefont {M.~L.}\ \bibnamefont {Peerlings}}, \bibinfo {author} {\bibfnamefont {B.~B.}\ \bibnamefont {Salzmann}},\ and\ \bibinfo {author} {\bibfnamefont {A.}~\bibnamefont {Meijerink}},\ }\bibfield  {title} {\bibinfo {title} {Luminescence line broadening of cdse nanoplatelets and quantum dots for application in w-leds},\ }\href@noop {} {\bibfield  {journal} {\bibinfo  {journal} {J. Phys. Chem. C}\ }\textbf {\bibinfo {volume} {124}},\ \bibinfo {pages} {12153} (\bibinfo {year} {2020})}\BibitemShut {NoStop}%
\bibitem [{\citenamefont {Tan}\ \emph {et~al.}(2019)\citenamefont {Tan}, \citenamefont {Kaewuam}, \citenamefont {Arnold}, \citenamefont {Chanu}, \citenamefont {Zhang}, \citenamefont {Safronova},\ and\ \citenamefont {Barrett}}]{10}%
  \BibitemOpen
  \bibfield  {author} {\bibinfo {author} {\bibfnamefont {T.~R.}\ \bibnamefont {Tan}}, \bibinfo {author} {\bibfnamefont {R.}~\bibnamefont {Kaewuam}}, \bibinfo {author} {\bibfnamefont {K.~J.}\ \bibnamefont {Arnold}}, \bibinfo {author} {\bibfnamefont {S.~R.}\ \bibnamefont {Chanu}}, \bibinfo {author} {\bibfnamefont {Z.}~\bibnamefont {Zhang}}, \bibinfo {author} {\bibfnamefont {M.}~\bibnamefont {Safronova}},\ and\ \bibinfo {author} {\bibfnamefont {M.~D.}\ \bibnamefont {Barrett}},\ }\bibfield  {title} {\bibinfo {title} {Suppressing inhomogeneous broadening in a lutetium multi-ion optical clock},\ }\href@noop {} {\bibfield  {journal} {\bibinfo  {journal} {Phys. Rev. Lett.}\ }\textbf {\bibinfo {volume} {123}},\ \bibinfo {pages} {063201} (\bibinfo {year} {2019})}\BibitemShut {NoStop}%
\bibitem [{\citenamefont {Fan}\ \emph {et~al.}(2019)\citenamefont {Fan}, \citenamefont {Kagalwala}, \citenamefont {Polyakov}, \citenamefont {Migdall},\ and\ \citenamefont {Goldschmidt}}]{11}%
  \BibitemOpen
  \bibfield  {author} {\bibinfo {author} {\bibfnamefont {H.}~\bibnamefont {Fan}}, \bibinfo {author} {\bibfnamefont {K.~H.}\ \bibnamefont {Kagalwala}}, \bibinfo {author} {\bibfnamefont {S.~V.}\ \bibnamefont {Polyakov}}, \bibinfo {author} {\bibfnamefont {A.~L.}\ \bibnamefont {Migdall}},\ and\ \bibinfo {author} {\bibfnamefont {E.~A.}\ \bibnamefont {Goldschmidt}},\ }\bibfield  {title} {\bibinfo {title} {Electromagnetically induced transparency in inhomogeneously broadened solid media},\ }\href@noop {} {\bibfield  {journal} {\bibinfo  {journal} {Phys. Rev. A}\ }\textbf {\bibinfo {volume} {99}},\ \bibinfo {pages} {053821} (\bibinfo {year} {2019})}\BibitemShut {NoStop}%
\bibitem [{\citenamefont {Andrejić}\ and\ \citenamefont {Pálffy}(2021)}]{12}%
  \BibitemOpen
  \bibfield  {author} {\bibinfo {author} {\bibfnamefont {P.}~\bibnamefont {Andrejić}}\ and\ \bibinfo {author} {\bibfnamefont {A.}~\bibnamefont {Pálffy}},\ }\bibfield  {title} {\bibinfo {title} {Superradiance and anomalous hyperfine splitting in inhomogeneous ensembles},\ }\href@noop {} {\bibfield  {journal} {\bibinfo  {journal} {Phys. Rev. A}\ }\textbf {\bibinfo {volume} {104}},\ \bibinfo {pages} {033702} (\bibinfo {year} {2021})}\BibitemShut {NoStop}%
\bibitem [{\citenamefont {Abishev}\ \emph {et~al.}(2019)\citenamefont {Abishev}, \citenamefont {Baibekov}, \citenamefont {Malkin}, \citenamefont {Popova}, \citenamefont {Pytalev},\ and\ \citenamefont {Klimin}}]{13}%
  \BibitemOpen
  \bibfield  {author} {\bibinfo {author} {\bibfnamefont {N.~M.}\ \bibnamefont {Abishev}}, \bibinfo {author} {\bibfnamefont {E.~I.}\ \bibnamefont {Baibekov}}, \bibinfo {author} {\bibfnamefont {B.~Z.}\ \bibnamefont {Malkin}}, \bibinfo {author} {\bibfnamefont {M.~N.}\ \bibnamefont {Popova}}, \bibinfo {author} {\bibfnamefont {D.~S.}\ \bibnamefont {Pytalev}},\ and\ \bibinfo {author} {\bibfnamefont {S.~A.}\ \bibnamefont {Klimin}},\ }\bibfield  {title} {\bibinfo {title} {Deformation broadening and the fine structure of spectral lines in optical spectra of dielectric crystals containing rare-earth ions},\ }\href@noop {} {\bibfield  {journal} {\bibinfo  {journal} {Phys. Solid State}\ }\textbf {\bibinfo {volume} {61}},\ \bibinfo {pages} {795} (\bibinfo {year} {2019})}\BibitemShut {NoStop}%
\bibitem [{\citenamefont {Petrusevich}\ \emph {et~al.}(2023)\citenamefont {Petrusevich}, \citenamefont {Bousquet}, \citenamefont {Osmialowski}, \citenamefont {Jacquemin}, \citenamefont {Luis},\ and\ \citenamefont {Zalesny}}]{14}%
  \BibitemOpen
  \bibfield  {author} {\bibinfo {author} {\bibfnamefont {E.~F.}\ \bibnamefont {Petrusevich}}, \bibinfo {author} {\bibfnamefont {M.~H.}\ \bibnamefont {Bousquet}}, \bibinfo {author} {\bibfnamefont {B.}~\bibnamefont {Osmialowski}}, \bibinfo {author} {\bibfnamefont {D.}~\bibnamefont {Jacquemin}}, \bibinfo {author} {\bibfnamefont {J.~M.}\ \bibnamefont {Luis}},\ and\ \bibinfo {author} {\bibfnamefont {R.}~\bibnamefont {Zalesny}},\ }\bibfield  {title} {\bibinfo {title} {Cost-effective simulations of vibrationally-resolved absorption spectra of fluorophores with machine-learning-based inhomogeneous broadening},\ }\href@noop {} {\bibfield  {journal} {\bibinfo  {journal} {J. Chem. Theory Comput.}\ }\textbf {\bibinfo {volume} {19}},\ \bibinfo {pages} {2304} (\bibinfo {year} {2023})}\BibitemShut {NoStop}%
\bibitem [{\citenamefont {Cerezo}\ \emph {et~al.}(2015)\citenamefont {Cerezo}, \citenamefont {Avila~Ferrer}, \citenamefont {Prampolini},\ and\ \citenamefont {Santoro}}]{15}%
  \BibitemOpen
  \bibfield  {author} {\bibinfo {author} {\bibfnamefont {J.}~\bibnamefont {Cerezo}}, \bibinfo {author} {\bibfnamefont {F.~J.}\ \bibnamefont {Avila~Ferrer}}, \bibinfo {author} {\bibfnamefont {G.}~\bibnamefont {Prampolini}},\ and\ \bibinfo {author} {\bibfnamefont {F.}~\bibnamefont {Santoro}},\ }\bibfield  {title} {\bibinfo {title} {Modeling solvent broadening on the vibronic spectra of a series of coumarin dyes. from implicit to explicit solvent models},\ }\href@noop {} {\bibfield  {journal} {\bibinfo  {journal} {J. Chem. Theory Comput.}\ }\textbf {\bibinfo {volume} {11}},\ \bibinfo {pages} {5810} (\bibinfo {year} {2015})}\BibitemShut {NoStop}%
\bibitem [{\citenamefont {Ferrer}\ \emph {et~al.}(2011)\citenamefont {Ferrer}, \citenamefont {Improta}, \citenamefont {Santoro},\ and\ \citenamefont {Barone}}]{16}%
  \BibitemOpen
  \bibfield  {author} {\bibinfo {author} {\bibfnamefont {F.~J.~A.}\ \bibnamefont {Ferrer}}, \bibinfo {author} {\bibfnamefont {R.}~\bibnamefont {Improta}}, \bibinfo {author} {\bibfnamefont {F.}~\bibnamefont {Santoro}},\ and\ \bibinfo {author} {\bibfnamefont {V.}~\bibnamefont {Barone}},\ }\bibfield  {title} {\bibinfo {title} {Computing the inhomogeneous broadening of electronic transitions in solution: A first-principle quantum mechanical approach},\ }\href@noop {} {\bibfield  {journal} {\bibinfo  {journal} {Phys. Chem. Chem. Phys.}\ }\textbf {\bibinfo {volume} {13}},\ \bibinfo {pages} {17007} (\bibinfo {year} {2011})}\BibitemShut {NoStop}%
\bibitem [{\citenamefont {Idrees}\ \emph {et~al.}(2020)\citenamefont {Idrees}, \citenamefont {Bacha}, \citenamefont {Khan}, \citenamefont {Ullah},\ and\ \citenamefont {Haneef}}]{17}%
  \BibitemOpen
  \bibfield  {author} {\bibinfo {author} {\bibfnamefont {M.}~\bibnamefont {Idrees}}, \bibinfo {author} {\bibfnamefont {B.~A.}\ \bibnamefont {Bacha}}, \bibinfo {author} {\bibfnamefont {H.}~\bibnamefont {Khan}}, \bibinfo {author} {\bibfnamefont {A.}~\bibnamefont {Ullah}},\ and\ \bibinfo {author} {\bibfnamefont {M.}~\bibnamefont {Haneef}},\ }\bibfield  {title} {\bibinfo {title} {Optical soliton through induced cesium doppler broadening medium},\ }\href@noop {} {\bibfield  {journal} {\bibinfo  {journal} {Phys. Scr.}\ }\textbf {\bibinfo {volume} {95}},\ \bibinfo {pages} {085102} (\bibinfo {year} {2020})}\BibitemShut {NoStop}%
\bibitem [{\citenamefont {Ismail}\ \emph {et~al.}(2018)\citenamefont {Ismail}, \citenamefont {Mohammed},\ and\ \citenamefont {Fouad}}]{18}%
  \BibitemOpen
  \bibfield  {author} {\bibinfo {author} {\bibfnamefont {A.}~\bibnamefont {Ismail}}, \bibinfo {author} {\bibfnamefont {M.}~\bibnamefont {Mohammed}},\ and\ \bibinfo {author} {\bibfnamefont {S.}~\bibnamefont {Fouad}},\ }\bibfield  {title} {\bibinfo {title} {Optical and structural properties of polyvinylidene fluoride (pvdf)/reduced graphene oxide (rgo) nanocomposites},\ }\href@noop {} {\bibfield  {journal} {\bibinfo  {journal} {J. Mol. Struct.}\ }\textbf {\bibinfo {volume} {1170}},\ \bibinfo {pages} {51} (\bibinfo {year} {2018})}\BibitemShut {NoStop}%
\bibitem [{\citenamefont {Hoffmann}\ \emph {et~al.}(2010)\citenamefont {Hoffmann}, \citenamefont {Bassler},\ and\ \citenamefont {Kohler}}]{19}%
  \BibitemOpen
  \bibfield  {author} {\bibinfo {author} {\bibfnamefont {S.~T.}\ \bibnamefont {Hoffmann}}, \bibinfo {author} {\bibfnamefont {H.}~\bibnamefont {Bassler}},\ and\ \bibinfo {author} {\bibfnamefont {A.}~\bibnamefont {Kohler}},\ }\bibfield  {title} {\bibinfo {title} {What determines inhomogeneous broadening of electronic transitions in conjugated polymers?},\ }\href@noop {} {\bibfield  {journal} {\bibinfo  {journal} {J. Phys. Chem. B}\ }\textbf {\bibinfo {volume} {114}},\ \bibinfo {pages} {17037} (\bibinfo {year} {2010})}\BibitemShut {NoStop}%
\bibitem [{\citenamefont {Reimer}\ \emph {et~al.}(2016)\citenamefont {Reimer}, \citenamefont {Bulgarini}, \citenamefont {Fognini}, \citenamefont {Heeres}, \citenamefont {Witek}, \citenamefont {Versteegh}, \citenamefont {Rubino}, \citenamefont {Braun}, \citenamefont {Kamp},\ and\ \citenamefont {Höfling}}]{20}%
  \BibitemOpen
  \bibfield  {author} {\bibinfo {author} {\bibfnamefont {M.~E.}\ \bibnamefont {Reimer}}, \bibinfo {author} {\bibfnamefont {G.}~\bibnamefont {Bulgarini}}, \bibinfo {author} {\bibfnamefont {A.}~\bibnamefont {Fognini}}, \bibinfo {author} {\bibfnamefont {R.~W.}\ \bibnamefont {Heeres}}, \bibinfo {author} {\bibfnamefont {B.~J.}\ \bibnamefont {Witek}}, \bibinfo {author} {\bibfnamefont {M.~A.}\ \bibnamefont {Versteegh}}, \bibinfo {author} {\bibfnamefont {A.}~\bibnamefont {Rubino}}, \bibinfo {author} {\bibfnamefont {T.}~\bibnamefont {Braun}}, \bibinfo {author} {\bibfnamefont {M.}~\bibnamefont {Kamp}},\ and\ \bibinfo {author} {\bibfnamefont {S.}~\bibnamefont {Höfling}},\ }\bibfield  {title} {\bibinfo {title} {Overcoming power broadening of the quantum dot emission in a pure wurtzite nanowire},\ }\href@noop {} {\bibfield  {journal} {\bibinfo  {journal} {Phys. Rev. B}\ }\textbf {\bibinfo {volume} {93}},\ \bibinfo {pages} {195316} (\bibinfo {year} {2016})}\BibitemShut {NoStop}%
\bibitem [{\citenamefont {Petrov}(2012)}]{petrov2012sums}%
  \BibitemOpen
  \bibfield  {author} {\bibinfo {author} {\bibfnamefont {V.~V.}\ \bibnamefont {Petrov}},\ }\href@noop {} {\emph {\bibinfo {title} {Sums of independent random variables}}},\ Vol.~\bibinfo {volume} {82}\ (\bibinfo  {publisher} {Springer Science \& Business Media},\ \bibinfo {year} {2012})\BibitemShut {NoStop}%
\bibitem [{\citenamefont {Zentile}\ \emph {et~al.}(2015)\citenamefont {Zentile}, \citenamefont {Keaveney}, \citenamefont {Mathew}, \citenamefont {Whiting}, \citenamefont {Adams},\ and\ \citenamefont {Hughes}}]{21}%
  \BibitemOpen
  \bibfield  {author} {\bibinfo {author} {\bibfnamefont {M.~A.}\ \bibnamefont {Zentile}}, \bibinfo {author} {\bibfnamefont {J.}~\bibnamefont {Keaveney}}, \bibinfo {author} {\bibfnamefont {R.~S.}\ \bibnamefont {Mathew}}, \bibinfo {author} {\bibfnamefont {D.~J.}\ \bibnamefont {Whiting}}, \bibinfo {author} {\bibfnamefont {C.~S.}\ \bibnamefont {Adams}},\ and\ \bibinfo {author} {\bibfnamefont {I.~G.}\ \bibnamefont {Hughes}},\ }\bibfield  {title} {\bibinfo {title} {Optimization of atomic faraday filters in the presence of homogeneous line broadening},\ }\href@noop {} {\bibfield  {journal} {\bibinfo  {journal} {J. Phys. B}\ }\textbf {\bibinfo {volume} {48}},\ \bibinfo {pages} {185001} (\bibinfo {year} {2015})}\BibitemShut {NoStop}%
\bibitem [{\citenamefont {Sun}\ \emph {et~al.}(2019)\citenamefont {Sun}, \citenamefont {Chernyak}, \citenamefont {Piryatinski},\ and\ \citenamefont {Sinitsyn}}]{22}%
  \BibitemOpen
  \bibfield  {author} {\bibinfo {author} {\bibfnamefont {C.}~\bibnamefont {Sun}}, \bibinfo {author} {\bibfnamefont {V.~Y.}\ \bibnamefont {Chernyak}}, \bibinfo {author} {\bibfnamefont {A.}~\bibnamefont {Piryatinski}},\ and\ \bibinfo {author} {\bibfnamefont {N.~A.}\ \bibnamefont {Sinitsyn}},\ }\bibfield  {title} {\bibinfo {title} {Cooperative light emission in the presence of strong inhomogeneous broadening},\ }\href@noop {} {\bibfield  {journal} {\bibinfo  {journal} {Phys. Rev. Lett.}\ }\textbf {\bibinfo {volume} {123}},\ \bibinfo {pages} {123605} (\bibinfo {year} {2019})}\BibitemShut {NoStop}%
\bibitem [{\citenamefont {Debnath}\ \emph {et~al.}(2019)\citenamefont {Debnath}, \citenamefont {Zhang},\ and\ \citenamefont {Mølmer}}]{23}%
  \BibitemOpen
  \bibfield  {author} {\bibinfo {author} {\bibfnamefont {K.}~\bibnamefont {Debnath}}, \bibinfo {author} {\bibfnamefont {Y.}~\bibnamefont {Zhang}},\ and\ \bibinfo {author} {\bibfnamefont {K.}~\bibnamefont {Mølmer}},\ }\bibfield  {title} {\bibinfo {title} {Collective dynamics of inhomogeneously broadened emitters coupled to an optical cavity with narrow linewidth},\ }\href@noop {} {\bibfield  {journal} {\bibinfo  {journal} {Phys. Rev. A}\ }\textbf {\bibinfo {volume} {100}},\ \bibinfo {pages} {053821} (\bibinfo {year} {2019})}\BibitemShut {NoStop}%
\bibitem [{\citenamefont {Houdre}\ \emph {et~al.}(1996)\citenamefont {Houdre}, \citenamefont {Stanley},\ and\ \citenamefont {Ilegems}}]{24}%
  \BibitemOpen
  \bibfield  {author} {\bibinfo {author} {\bibfnamefont {R.}~\bibnamefont {Houdre}}, \bibinfo {author} {\bibfnamefont {R.}~\bibnamefont {Stanley}},\ and\ \bibinfo {author} {\bibfnamefont {M.}~\bibnamefont {Ilegems}},\ }\bibfield  {title} {\bibinfo {title} {Vacuum-field rabi splitting in the presence of inhomogeneous broadening: Resolution of a homogeneous linewidth in an inhomogeneously broadened system},\ }\href@noop {} {\bibfield  {journal} {\bibinfo  {journal} {Phys. Rev. A}\ }\textbf {\bibinfo {volume} {53}},\ \bibinfo {pages} {2711} (\bibinfo {year} {1996})}\BibitemShut {NoStop}%
\bibitem [{\citenamefont {Diniz}\ \emph {et~al.}(2011)\citenamefont {Diniz}, \citenamefont {Portolan}, \citenamefont {Ferreira}, \citenamefont {Gerard}, \citenamefont {Bertet},\ and\ \citenamefont {Auffeves}}]{25}%
  \BibitemOpen
  \bibfield  {author} {\bibinfo {author} {\bibfnamefont {I.}~\bibnamefont {Diniz}}, \bibinfo {author} {\bibfnamefont {S.}~\bibnamefont {Portolan}}, \bibinfo {author} {\bibfnamefont {R.}~\bibnamefont {Ferreira}}, \bibinfo {author} {\bibfnamefont {J.}~\bibnamefont {Gerard}}, \bibinfo {author} {\bibfnamefont {P.}~\bibnamefont {Bertet}},\ and\ \bibinfo {author} {\bibfnamefont {A.}~\bibnamefont {Auffeves}},\ }\bibfield  {title} {\bibinfo {title} {Strongly coupling a cavity to inhomogeneous ensembles of emitters: Potential for long-lived solid-state quantum memories},\ }\href@noop {} {\bibfield  {journal} {\bibinfo  {journal} {Phys. Rev. A}\ }\textbf {\bibinfo {volume} {84}},\ \bibinfo {pages} {063810} (\bibinfo {year} {2011})}\BibitemShut {NoStop}%
\bibitem [{\citenamefont {Mermilliod}\ \emph {et~al.}(1992)\citenamefont {Mermilliod}, \citenamefont {Romero}, \citenamefont {Chartier}, \citenamefont {Garapon},\ and\ \citenamefont {Moncorgé}}]{26}%
  \BibitemOpen
  \bibfield  {author} {\bibinfo {author} {\bibfnamefont {N.}~\bibnamefont {Mermilliod}}, \bibinfo {author} {\bibfnamefont {R.}~\bibnamefont {Romero}}, \bibinfo {author} {\bibfnamefont {I.}~\bibnamefont {Chartier}}, \bibinfo {author} {\bibfnamefont {C.}~\bibnamefont {Garapon}},\ and\ \bibinfo {author} {\bibfnamefont {R.}~\bibnamefont {Moncorgé}},\ }\bibfield  {title} {\bibinfo {title} {Performance of various diode-pumped nd: laser materials: influence of inhomogeneous broadening},\ }\href@noop {} {\bibfield  {journal} {\bibinfo  {journal} {IEEE J. Quantum Electron.}\ }\textbf {\bibinfo {volume} {28}},\ \bibinfo {pages} {1179} (\bibinfo {year} {1992})}\BibitemShut {NoStop}%
\bibitem [{\citenamefont {Andreani}\ \emph {et~al.}(1998)\citenamefont {Andreani}, \citenamefont {Panzarini}, \citenamefont {Kavokin},\ and\ \citenamefont {Vladimirova}}]{27}%
  \BibitemOpen
  \bibfield  {author} {\bibinfo {author} {\bibfnamefont {L.~C.}\ \bibnamefont {Andreani}}, \bibinfo {author} {\bibfnamefont {G.}~\bibnamefont {Panzarini}}, \bibinfo {author} {\bibfnamefont {A.~V.}\ \bibnamefont {Kavokin}},\ and\ \bibinfo {author} {\bibfnamefont {M.~R.}\ \bibnamefont {Vladimirova}},\ }\bibfield  {title} {\bibinfo {title} {Effect of inhomogeneous broadening on optical properties of excitons in quantum wells},\ }\href@noop {} {\bibfield  {journal} {\bibinfo  {journal} {Phys. Rev. B}\ }\textbf {\bibinfo {volume} {57}},\ \bibinfo {pages} {4670} (\bibinfo {year} {1998})}\BibitemShut {NoStop}%
\bibitem [{\citenamefont {Zyablovsky}\ \emph {et~al.}(2021)\citenamefont {Zyablovsky}, \citenamefont {Doronin}, \citenamefont {Andrianov}, \citenamefont {Pukhov}, \citenamefont {Lozovik}, \citenamefont {Vinogradov},\ and\ \citenamefont {Lisyansky}}]{33}%
  \BibitemOpen
  \bibfield  {author} {\bibinfo {author} {\bibfnamefont {A.~A.}\ \bibnamefont {Zyablovsky}}, \bibinfo {author} {\bibfnamefont {I.~V.}\ \bibnamefont {Doronin}}, \bibinfo {author} {\bibfnamefont {E.~S.}\ \bibnamefont {Andrianov}}, \bibinfo {author} {\bibfnamefont {A.~A.}\ \bibnamefont {Pukhov}}, \bibinfo {author} {\bibfnamefont {Y.~E.}\ \bibnamefont {Lozovik}}, \bibinfo {author} {\bibfnamefont {A.~P.}\ \bibnamefont {Vinogradov}},\ and\ \bibinfo {author} {\bibfnamefont {A.~A.}\ \bibnamefont {Lisyansky}},\ }\bibfield  {title} {\bibinfo {title} {Exceptional points as lasing prethresholds},\ }\href@noop {} {\bibfield  {journal} {\bibinfo  {journal} {Laser Photonics Rev.}\ }\textbf {\bibinfo {volume} {15}},\ \bibinfo {pages} {2000450} (\bibinfo {year} {2021})}\BibitemShut {NoStop}%
\bibitem [{\citenamefont {Shank}(1975)}]{shank1975physics}%
  \BibitemOpen
  \bibfield  {author} {\bibinfo {author} {\bibfnamefont {C.~V.}\ \bibnamefont {Shank}},\ }\bibfield  {title} {\bibinfo {title} {Physics of dye lasers},\ }\href@noop {} {\bibfield  {journal} {\bibinfo  {journal} {Rev. Mod. Phys.}\ }\textbf {\bibinfo {volume} {47}},\ \bibinfo {pages} {649} (\bibinfo {year} {1975})}\BibitemShut {NoStop}%
\bibitem [{\citenamefont {Zrenner}\ \emph {et~al.}(2002)\citenamefont {Zrenner}, \citenamefont {Beham}, \citenamefont {Stufler}, \citenamefont {Findeis}, \citenamefont {Bichler},\ and\ \citenamefont {Abstreiter}}]{zrenner2002coherent}%
  \BibitemOpen
  \bibfield  {author} {\bibinfo {author} {\bibfnamefont {A.}~\bibnamefont {Zrenner}}, \bibinfo {author} {\bibfnamefont {E.}~\bibnamefont {Beham}}, \bibinfo {author} {\bibfnamefont {S.}~\bibnamefont {Stufler}}, \bibinfo {author} {\bibfnamefont {F.}~\bibnamefont {Findeis}}, \bibinfo {author} {\bibfnamefont {M.}~\bibnamefont {Bichler}},\ and\ \bibinfo {author} {\bibfnamefont {G.}~\bibnamefont {Abstreiter}},\ }\bibfield  {title} {\bibinfo {title} {Coherent properties of a two-level system based on a quantum-dot photodiode},\ }\href@noop {} {\bibfield  {journal} {\bibinfo  {journal} {Nature}\ }\textbf {\bibinfo {volume} {418}},\ \bibinfo {pages} {612} (\bibinfo {year} {2002})}\BibitemShut {NoStop}%
\bibitem [{\citenamefont {He}\ \emph {et~al.}(2019)\citenamefont {He}, \citenamefont {Wang}, \citenamefont {Wang}, \citenamefont {Chen}, \citenamefont {Ding}, \citenamefont {Qin}, \citenamefont {Duan}, \citenamefont {Chen}, \citenamefont {Li}, \citenamefont {Liu} \emph {et~al.}}]{he2019coherently}%
  \BibitemOpen
  \bibfield  {author} {\bibinfo {author} {\bibfnamefont {Y.-M.}\ \bibnamefont {He}}, \bibinfo {author} {\bibfnamefont {H.}~\bibnamefont {Wang}}, \bibinfo {author} {\bibfnamefont {C.}~\bibnamefont {Wang}}, \bibinfo {author} {\bibfnamefont {M.-C.}\ \bibnamefont {Chen}}, \bibinfo {author} {\bibfnamefont {X.}~\bibnamefont {Ding}}, \bibinfo {author} {\bibfnamefont {J.}~\bibnamefont {Qin}}, \bibinfo {author} {\bibfnamefont {Z.-C.}\ \bibnamefont {Duan}}, \bibinfo {author} {\bibfnamefont {S.}~\bibnamefont {Chen}}, \bibinfo {author} {\bibfnamefont {J.-P.}\ \bibnamefont {Li}}, \bibinfo {author} {\bibfnamefont {R.-Z.}\ \bibnamefont {Liu}}, \emph {et~al.},\ }\bibfield  {title} {\bibinfo {title} {Coherently driving a single quantum two-level system with dichromatic laser pulses},\ }\href@noop {} {\bibfield  {journal} {\bibinfo  {journal} {Nat. Phys.}\ }\textbf {\bibinfo {volume} {15}},\ \bibinfo {pages} {941} (\bibinfo {year} {2019})}\BibitemShut {NoStop}%
\bibitem [{\citenamefont {Cartar}\ \emph {et~al.}(2017)\citenamefont {Cartar}, \citenamefont {M{\o}rk},\ and\ \citenamefont {Hughes}}]{cartar2017self}%
  \BibitemOpen
  \bibfield  {author} {\bibinfo {author} {\bibfnamefont {W.}~\bibnamefont {Cartar}}, \bibinfo {author} {\bibfnamefont {J.}~\bibnamefont {M{\o}rk}},\ and\ \bibinfo {author} {\bibfnamefont {S.}~\bibnamefont {Hughes}},\ }\bibfield  {title} {\bibinfo {title} {Self-consistent maxwell-bloch model of quantum-dot photonic-crystal-cavity lasers},\ }\href@noop {} {\bibfield  {journal} {\bibinfo  {journal} {Physical Review A}\ }\textbf {\bibinfo {volume} {96}},\ \bibinfo {pages} {023859} (\bibinfo {year} {2017})}\BibitemShut {NoStop}%
\bibitem [{\citenamefont {Meiser}\ \emph {et~al.}(2009)\citenamefont {Meiser}, \citenamefont {Ye}, \citenamefont {Carlson},\ and\ \citenamefont {Holland}}]{meiser2009prospects}%
  \BibitemOpen
  \bibfield  {author} {\bibinfo {author} {\bibfnamefont {D.}~\bibnamefont {Meiser}}, \bibinfo {author} {\bibfnamefont {J.}~\bibnamefont {Ye}}, \bibinfo {author} {\bibfnamefont {D.}~\bibnamefont {Carlson}},\ and\ \bibinfo {author} {\bibfnamefont {M.}~\bibnamefont {Holland}},\ }\bibfield  {title} {\bibinfo {title} {Prospects for a millihertz-linewidth laser},\ }\href@noop {} {\bibfield  {journal} {\bibinfo  {journal} {Phys. Rev. Lett.}\ }\textbf {\bibinfo {volume} {102}},\ \bibinfo {pages} {163601} (\bibinfo {year} {2009})}\BibitemShut {NoStop}%
\bibitem [{\citenamefont {Bohnet}\ \emph {et~al.}(2012)\citenamefont {Bohnet}, \citenamefont {Chen}, \citenamefont {Weiner}, \citenamefont {Meiser}, \citenamefont {Holland},\ and\ \citenamefont {Thompson}}]{bohnet2012steady}%
  \BibitemOpen
  \bibfield  {author} {\bibinfo {author} {\bibfnamefont {J.~G.}\ \bibnamefont {Bohnet}}, \bibinfo {author} {\bibfnamefont {Z.}~\bibnamefont {Chen}}, \bibinfo {author} {\bibfnamefont {J.~M.}\ \bibnamefont {Weiner}}, \bibinfo {author} {\bibfnamefont {D.}~\bibnamefont {Meiser}}, \bibinfo {author} {\bibfnamefont {M.~J.}\ \bibnamefont {Holland}},\ and\ \bibinfo {author} {\bibfnamefont {J.~K.}\ \bibnamefont {Thompson}},\ }\bibfield  {title} {\bibinfo {title} {A steady-state superradiant laser with less than one intracavity photon},\ }\href@noop {} {\bibfield  {journal} {\bibinfo  {journal} {Nature}\ }\textbf {\bibinfo {volume} {484}},\ \bibinfo {pages} {78} (\bibinfo {year} {2012})}\BibitemShut {NoStop}%
\bibitem [{\citenamefont {Scully}\ and\ \citenamefont {Zubairy}(1997)}]{29}%
  \BibitemOpen
  \bibfield  {author} {\bibinfo {author} {\bibfnamefont {M.~O.}\ \bibnamefont {Scully}}\ and\ \bibinfo {author} {\bibfnamefont {M.~S.}\ \bibnamefont {Zubairy}},\ }\href@noop {} {\emph {\bibinfo {title} {Quantum optics}}}\ (\bibinfo  {publisher} {Cambridge University Press},\ \bibinfo {address} {Cambridge},\ \bibinfo {year} {1997})\BibitemShut {NoStop}%
\bibitem [{\citenamefont {Siegman}(1986)}]{30}%
  \BibitemOpen
  \bibfield  {author} {\bibinfo {author} {\bibfnamefont {A.~E.}\ \bibnamefont {Siegman}},\ }\href@noop {} {\emph {\bibinfo {title} {Lasers}}}\ (\bibinfo  {publisher} {University Science Books},\ \bibinfo {address} {Mill Valley, CA},\ \bibinfo {year} {1986})\ p.\ \bibinfo {pages} {654}\BibitemShut {NoStop}%
\bibitem [{\citenamefont {Doronin}\ \emph {et~al.}(2019)\citenamefont {Doronin}, \citenamefont {Zyablovsky}, \citenamefont {Andrianov}, \citenamefont {Pukhov},\ and\ \citenamefont {Vinogradov}}]{31}%
  \BibitemOpen
  \bibfield  {author} {\bibinfo {author} {\bibfnamefont {I.~V.}\ \bibnamefont {Doronin}}, \bibinfo {author} {\bibfnamefont {A.~A.}\ \bibnamefont {Zyablovsky}}, \bibinfo {author} {\bibfnamefont {E.~S.}\ \bibnamefont {Andrianov}}, \bibinfo {author} {\bibfnamefont {A.~A.}\ \bibnamefont {Pukhov}},\ and\ \bibinfo {author} {\bibfnamefont {A.~P.}\ \bibnamefont {Vinogradov}},\ }\bibfield  {title} {\bibinfo {title} {Lasing without inversion due to parametric instability of the laser near the exceptional point},\ }\href@noop {} {\bibfield  {journal} {\bibinfo  {journal} {Phys. Rev. A}\ }\textbf {\bibinfo {volume} {100}},\ \bibinfo {pages} {021801} (\bibinfo {year} {2019})}\BibitemShut {NoStop}%
\bibitem [{\citenamefont {Doronin}\ \emph {et~al.}(2021{\natexlab{a}})\citenamefont {Doronin}, \citenamefont {Zyablovsky},\ and\ \citenamefont {Andrianov}}]{32}%
  \BibitemOpen
  \bibfield  {author} {\bibinfo {author} {\bibfnamefont {I.~V.}\ \bibnamefont {Doronin}}, \bibinfo {author} {\bibfnamefont {A.~A.}\ \bibnamefont {Zyablovsky}},\ and\ \bibinfo {author} {\bibfnamefont {E.~S.}\ \bibnamefont {Andrianov}},\ }\bibfield  {title} {\bibinfo {title} {Strong-coupling-assisted formation of coherent radiation below the lasing threshold},\ }\href@noop {} {\bibfield  {journal} {\bibinfo  {journal} {Opt. Express}\ }\textbf {\bibinfo {volume} {29}},\ \bibinfo {pages} {5624} (\bibinfo {year} {2021}{\natexlab{a}})}\BibitemShut {NoStop}%
\bibitem [{\citenamefont {Zyablovsky}\ \emph {et~al.}(2017)\citenamefont {Zyablovsky}, \citenamefont {Andrianov}, \citenamefont {Nechepurenko}, \citenamefont {Dorofeenko}, \citenamefont {Pukhov},\ and\ \citenamefont {Vinogradov}}]{34}%
  \BibitemOpen
  \bibfield  {author} {\bibinfo {author} {\bibfnamefont {A.~A.}\ \bibnamefont {Zyablovsky}}, \bibinfo {author} {\bibfnamefont {E.~S.}\ \bibnamefont {Andrianov}}, \bibinfo {author} {\bibfnamefont {I.~A.}\ \bibnamefont {Nechepurenko}}, \bibinfo {author} {\bibfnamefont {A.~V.}\ \bibnamefont {Dorofeenko}}, \bibinfo {author} {\bibfnamefont {A.~A.}\ \bibnamefont {Pukhov}},\ and\ \bibinfo {author} {\bibfnamefont {A.~P.}\ \bibnamefont {Vinogradov}},\ }\bibfield  {title} {\bibinfo {title} {Approach for describing spatial dynamics of quantum light-matter interaction in dispersive dissipative media},\ }\href@noop {} {\bibfield  {journal} {\bibinfo  {journal} {Phys. Rev. A}\ }\textbf {\bibinfo {volume} {95}},\ \bibinfo {pages} {053835} (\bibinfo {year} {2017})}\BibitemShut {NoStop}%
\bibitem [{\citenamefont {Doronin}\ \emph {et~al.}(2021{\natexlab{b}})\citenamefont {Doronin}, \citenamefont {Zyablovsky}, \citenamefont {Andrianov}, \citenamefont {Pukhov}, \citenamefont {Lozovik},\ and\ \citenamefont {Vinogradov}}]{doronin2021universal}%
  \BibitemOpen
  \bibfield  {author} {\bibinfo {author} {\bibfnamefont {I.~V.}\ \bibnamefont {Doronin}}, \bibinfo {author} {\bibfnamefont {A.~A.}\ \bibnamefont {Zyablovsky}}, \bibinfo {author} {\bibfnamefont {E.~S.}\ \bibnamefont {Andrianov}}, \bibinfo {author} {\bibfnamefont {A.~A.}\ \bibnamefont {Pukhov}}, \bibinfo {author} {\bibfnamefont {Y.~E.}\ \bibnamefont {Lozovik}},\ and\ \bibinfo {author} {\bibfnamefont {A.~P.}\ \bibnamefont {Vinogradov}},\ }\bibfield  {title} {\bibinfo {title} {Universal lasing condition},\ }\href@noop {} {\bibfield  {journal} {\bibinfo  {journal} {Sci. Rep.}\ }\textbf {\bibinfo {volume} {11}},\ \bibinfo {pages} {4197} (\bibinfo {year} {2021}{\natexlab{b}})}\BibitemShut {NoStop}%
\bibitem [{\citenamefont {Abramowitz}\ and\ \citenamefont {Stegun}(1968)}]{abramowitz1968handbook}%
  \BibitemOpen
  \bibfield  {author} {\bibinfo {author} {\bibfnamefont {M.}~\bibnamefont {Abramowitz}}\ and\ \bibinfo {author} {\bibfnamefont {I.~A.}\ \bibnamefont {Stegun}},\ }\href@noop {} {\emph {\bibinfo {title} {Handbook of mathematical functions with formulas, graphs, and mathematical tables}}},\ Vol.~\bibinfo {volume} {55}\ (\bibinfo  {publisher} {US Government printing office},\ \bibinfo {year} {1968})\BibitemShut {NoStop}%
\bibitem [{\citenamefont {Anappara}\ \emph {et~al.}(2009)\citenamefont {Anappara}, \citenamefont {De~Liberato}, \citenamefont {Tredicucci}, \citenamefont {Ciuti}, \citenamefont {Biasiol}, \citenamefont {Sorba},\ and\ \citenamefont {Beltram}}]{28}%
  \BibitemOpen
  \bibfield  {author} {\bibinfo {author} {\bibfnamefont {A.~A.}\ \bibnamefont {Anappara}}, \bibinfo {author} {\bibfnamefont {S.}~\bibnamefont {De~Liberato}}, \bibinfo {author} {\bibfnamefont {A.}~\bibnamefont {Tredicucci}}, \bibinfo {author} {\bibfnamefont {C.}~\bibnamefont {Ciuti}}, \bibinfo {author} {\bibfnamefont {G.}~\bibnamefont {Biasiol}}, \bibinfo {author} {\bibfnamefont {L.}~\bibnamefont {Sorba}},\ and\ \bibinfo {author} {\bibfnamefont {F.}~\bibnamefont {Beltram}},\ }\bibfield  {title} {\bibinfo {title} {Signatures of the ultrastrong light-matter coupling regime},\ }\href@noop {} {\bibfield  {journal} {\bibinfo  {journal} {Phys. Rev. B}\ }\textbf {\bibinfo {volume} {79}},\ \bibinfo {pages} {201303} (\bibinfo {year} {2009})}\BibitemShut {NoStop}%
\end{thebibliography}%

\end{document}